\documentclass[twocolumn, tighten, times]{aastex61}
\usepackage{natbib}
\newcommand{\thisstar}{HD\,141569\,A\ }

\received{September 22, 2016}
\revised{November 2, 2016}
\accepted{November 16, 2016}

\submitjournal{AJ}

\shorttitle{The inner disk of \thisstar}
\shortauthors{Mawet et al.}

\begin{document}

\title{Characterization of the inner disk around \thisstar\\ from Keck/NIRC2 L-band vortex coronagraphy}

\correspondingauthor{Dimitri Mawet}
\email{dmawet@astro.caltech.edu}

\author{Dimitri Mawet}
\affiliation{Department of Astronomy, California Institute of Technology, 1200 E. California Blvd, MC 249-17, Pasadena, CA 91125 USA}
\affiliation{Jet Propulsion Laboratory, California Institute of Technology, 4800 Oak Grove Drive, Pasadena, CA 91109}

\author{\'Elodie Choquet}
\altaffiliation{Hubble fellow.}
\affiliation{Jet Propulsion Laboratory, California Institute of Technology, 4800 Oak Grove Drive, Pasadena, CA 91109}

\author{Olivier Absil}
\altaffiliation{F.R.S.-FNRS Research Associate.}
\affiliation{Space sciences, Technologies and Astrophysics Research (STAR) Institute, Universit\'e de Li\`ege, 19 All\'ee du Six Ao\^ut, B-4000 Li\`ege, Belgium}

\author{Elsa Huby}
\altaffiliation{F.R.S.-FNRS Postdoctoral Researcher.}
\affiliation{Space sciences, Technologies and Astrophysics Research (STAR) Institute, Universit\'e de Li\`ege, 19 All\'ee du Six Ao\^ut, B-4000 Li\`ege, Belgium}

\author{Michael Bottom}
\affiliation{Department of Astronomy, California Institute of Technology, 1200 E. California Blvd, MC 249-17, Pasadena, CA 91125 USA}
\affiliation{Jet Propulsion Laboratory, California Institute of Technology, 4800 Oak Grove Drive, Pasadena, CA 91109}

\author{Eugene Serabyn}
\affiliation{Jet Propulsion Laboratory, California Institute of Technology, 4800 Oak Grove Drive, Pasadena, CA 91109}

\author{Bruno Femenia}
\affiliation{W. M. Keck Observatory, 65-1120 Mamalahoa Hwy., Kamuela, HI 96743, USA}

\author{J\'er\'emy Lebreton}
\affiliation{NASA Exoplanet Science Institute, California Institute of Technology, 770 South Wilson Avenue, Pasadena, CA 91125, USA}

\author{Keith Matthews}
\affiliation{Department of Astronomy, California Institute of Technology, 1200 E. California Blvd, MC 249-17, Pasadena, CA 91125 USA}

\author{Carlos A. Gomez Gonzalez}
\affiliation{Space sciences, Technologies and Astrophysics Research (STAR) Institute, Universit\'e de Li\`ege, 19 All\'ee du Six Ao\^ut, B-4000 Li\`ege, Belgium}

\author{Olivier Wertz}
\affiliation{Argelander-Institut f\"ur Astronomie, Auf dem H\"ugel 71, D-53121 Bonn, Germany}
\affiliation{Space sciences, Technologies and Astrophysics Research (STAR) Institute, Universit\'e de Li\`ege, 19 All\'ee du Six Ao\^ut, B-4000 Li\`ege, Belgium}

\author{Brunella Carlomagno}
\affiliation{Space sciences, Technologies and Astrophysics Research (STAR) Institute, Universit\'e de Li\`ege, 19 All\'ee du Six Ao\^ut, B-4000 Li\`ege, Belgium}

\author{Valentin Christiaens}
\affiliation{Space sciences, Technologies and Astrophysics Research (STAR) Institute, Universit\'e de Li\`ege, 19 All\'ee du Six Ao\^ut, B-4000 Li\`ege, Belgium}
\affiliation{Departamento de Astronom\'ia, Universidad de Chile, Casilla 36-D, Santiago, Chile}

\author{Denis Defr\`ere}
\affiliation{Space sciences, Technologies and Astrophysics Research (STAR) Institute, Universit\'e de Li\`ege, 19 All\'ee du Six Ao\^ut, B-4000 Li\`ege, Belgium}
\affiliation{Steward Observatory, Department of Astronomy, University of Arizona, 933 N. Cherry Ave, Tucson, AZ 85721, USA}

\author{Christian Delacroix}
\affiliation{Space sciences, Technologies and Astrophysics Research (STAR) Institute, Universit\'e de Li\`ege, 19 All\'ee du Six Ao\^ut, B-4000 Li\`ege, Belgium}
\affiliation{Sibley School of Mechanical and Aerospace Engineering, Cornell University, Ithaca, NY 14853, USA}

\author{Pontus Forsberg}
\affiliation{Department of Engineering Sciences, {\AA}ngstr\"{o}m Laboratory, Uppsala University, Box 534, 751 21 Uppsala, Sweden}

\author{Serge Habraken}
\affiliation{Space sciences, Technologies and Astrophysics Research (STAR) Institute, Universit\'e de Li\`ege, 19 All\'ee du Six Ao\^ut, B-4000 Li\`ege, Belgium}

\author{Aissa Jolivet}
\affiliation{Space sciences, Technologies and Astrophysics Research (STAR) Institute, Universit\'e de Li\`ege, 19 All\'ee du Six Ao\^ut, B-4000 Li\`ege, Belgium}

\author{Mikael Karlsson}
\affiliation{Department of Engineering Sciences, {\AA}ngstr\"{o}m Laboratory, Uppsala University, Box 534, 751 21 Uppsala, Sweden}

\author{Julien Milli}
\affiliation{European Southern Observatory, Alonso de Cord\'ova 3107, Vitacura, Santiago, Chile}

\author{Christophe Pinte}
\affiliation{Univ. Grenoble Alpes, IPAG, F-38000 Grenoble, France CNRS, IPAG, F-38000 Grenoble, France}

\author{Pierre Piron}
\affiliation{Space sciences, Technologies and Astrophysics Research (STAR) Institute, Universit\'e de Li\`ege, 19 All\'ee du Six Ao\^ut, B-4000 Li\`ege, Belgium}
\affiliation{Department of Engineering Sciences, {\AA}ngstr\"{o}m Laboratory, Uppsala University, Box 534, 751 21 Uppsala, Sweden}

\author{Maddalena Reggiani}
\affiliation{Space sciences, Technologies and Astrophysics Research (STAR) Institute, Universit\'e de Li\`ege, 19 All\'ee du Six Ao\^ut, B-4000 Li\`ege, Belgium}

\author{Jean Surdej}
\affiliation{Space sciences, Technologies and Astrophysics Research (STAR) Institute, Universit\'e de Li\`ege, 19 All\'ee du Six Ao\^ut, B-4000 Li\`ege, Belgium}

\author{Ernesto Vargas Catalan}
\affiliation{Department of Engineering Sciences, {\AA}ngstr\"{o}m Laboratory, Uppsala University, Box 534, 751 21 Uppsala, Sweden}

\begin{abstract}
\thisstar is a pre-main sequence B9.5 Ve star surrounded by a prominent and complex circumstellar disk, likely still in a transition stage from protoplanetary to debris disk phase. Here, we present a new image of the third inner disk component of \thisstar made in the L' band (3.8 $\mu$m) during the commissioning of the vector vortex coronagraph recently installed in the near-infrared imager and spectrograph NIRC2 behind the W.M. Keck Observatory Keck II adaptive optics system. We used reference point spread function subtraction, which reveals the innermost disk component from the inner working distance of $\simeq 23$ AU and up to $\simeq 70$ AU. The spatial scale of our detection roughly corresponds to the optical and near-infrared scattered light, thermal Q, N and $8.6 \mu$m PAH emission reported earlier. We also see an outward progression in dust location from the L'-band to the H-band (VLT/SPHERE image) to the visible (HST/STIS image), likely indicative of dust blowout. The warm disk component is nested deep inside the two outer belts imaged by HST NICMOS in 1999 (respectively at 406 and 245 AU). We fit our new L'-band image and spectral energy distribution of \thisstar with the radiative transfer code MCFOST. Our best-fit models favor pure olivine grains, and are consistent with the composition of the outer belts. While our image shows a putative very-faint point-like clump or source embedded in the inner disk, we did not detect any true companion within the gap between the inner disk and the first outer ring, at a sensitivity of a few Jupiter masses. 
\end{abstract}


\section{Introduction}

The study of the morphology and composition of circumstellar disks in systems of different ages and masses allows us to probe different stages of the formation and evolution of planetary systems in diverse environmental conditions. This endeavor is complementary to the search of exoplanets, characterization of outstanding extra-solar planetary systems and planet population demographics. These synergistic approaches should allow us to relate the diversity of planetary systems observed by large-scale surveys to the initial conditions of stellar and planet formation, the ultimate goal of comparative exoplanetology.

\begin{figure}[!t]
  \centering
\includegraphics[width=8.5cm]{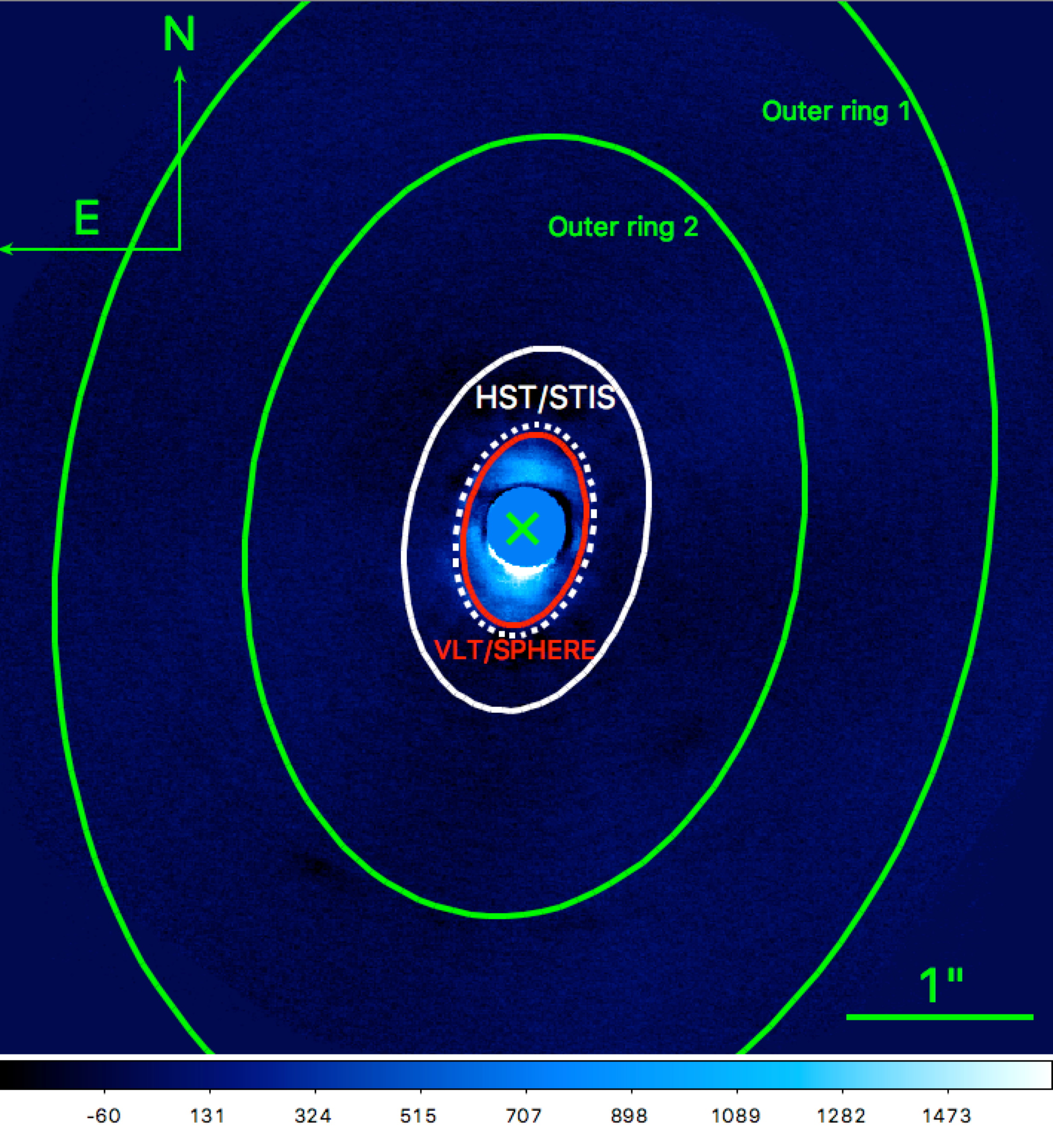}
  \caption{Wide field view of \thisstar from our L'-band vortex coronagraph data, showing the trace of the outer rings, not detected here. The inner white plain and dashed ellipses approximately delineate the disk component detected by HST/STIS \citep{2016arXiv160106560K}, clearly outside our detection. The red ellipse shows the approximate location of the main, brightest inner ring reported by \citet{Perrot:2016bt} using VLT/SPHERE. The color scale in the image is linear and in analog to digital units (ADU).
  \label{fig1}}
\end{figure}

\thisstar, a B9.5 Ve pre-main sequence star in a triple system $116\pm 8$ pc from the Sun \citep{2007A&A...474..653V}, is a natural testbed for stellar and planetary system evolution theories (Table \ref{tbl-1}). It is a young star with an age estimate of $5\pm 3$ Myr \citep{1999ApJ...525L..53W}, and is understood to be undergoing its transition from protoplanetary disk to debris disk. Using HST-NICMOS imaging, \citet{1999A&A...350L..51A} and \citet{1999ApJ...525L..53W} reported the discovery of two massive rings of dust around \thisstar almost simultaneously. The two resolved outer rings are roughly delineated in Fig.~\ref{fig1} by the green ellipses. \citet{2003AJ....126..385C} reported a fractional infrared excess luminosity of $L_{disk}/L_{star}=8.4\times 10^{-3}$, three times that of the $\beta$ Pictoris debris disk. \citet{2014A&A...561A..50T} reported a disk-averaged gas-to-dust ratio of $\simeq 100$, close to the initial interstellar value. All available empirical evidence clearly indicates that \thisstar is still transitioning from the protoplanetary gas-rich to the canonical gas-poor debris disk phase. 

The recent study of \thisstar with Gemini/NICI \citep{2015MNRAS.450.4446B} reported outer and inner ring radii of $406 \pm 13.3$ AU and $245\pm 3.0$, common position angle of $-11.3^\circ$ (resp. $-8.9^\circ$), and inclination of $43.7^\circ$ (resp. $44.9^\circ$). \citet{2015MNRAS.450.4446B} find an offset of 4 AU between the inner ring center and the stellar position, possibly hinting at the presence of unseen companion. This result is however disputed by a re-analysis of the same NICI data-set combined with HST/NICMOS data by \citet{2016ApJ...818..150M}, who found no significant offset. \citet{2015MNRAS.450.4446B} also report an additional arc-like feature between the inner and outer ring only evident on the East side, and claims an evacuated cavity from 175 AU inwards.

The discovery of the third, warm disk innermost component of \thisstar was first reported in \citet{Fisher2000} at 10.8 and 18.2 $\mu$m, using OSCIR on Keck. The mid-infrared emission from the warm disk was also confirmed later on by \citet{Marsh2002}, and was detected out to a radius of 100 AU, well inside the two outer belts (Fig.~\ref{fig1}) with a profile and brightness reminiscent of the debris disks around HR 4796A and $\beta$ Pictoris. \citet{Fisher2000} used the 10.8 and 18.2 $\mu$m images to place a lower limit of 170 K on the temperature and an upper limit of 2 $\mu$m on the diameter of the dust grains responsible for the mid-IR emission, assuming astronomical silicates. \citet{Fisher2000} argue that contamination of the mid-infrared flux by PAH emission marginally affects the derived grain size. \citet{2014A&A...561A..50T} discuss 2005 VISIR data obtained in the PAH filter at $8.6 \mu$m, and report the detection of infrared emission overlapping with the warm disk of \citet{Fisher2000} and \citet{Marsh2002}. The VISIR image of the warm inner disk component was interpreted as emission from PAHs by \citet{2014A&A...561A..50T}. The fitted disk-averaged gas-to-dust-mass ratio is about 100, with a gas mass of $4.9\times 10^{-4}M_{\odot}$ and accounts for multi-wavelength data, in particular Herschel PACS spectra. This result implies that if the disk was originally massive, the gas and the dust would have dissipated at the same rate.

\begin{deluxetable}{cc}
\tabletypesize{\scriptsize}

\tablecaption{Properties of \thisstar \label{tbl-1}}
\tablewidth{0pt}
\tablehead{
\colhead{Properties} & \colhead{value} 
}
\startdata
Coord. (hms, dms) & 15 49 57.74785 -03 55 16.3430 \\
Spectral type & B9.5 Ve \\
Distance (pc) & $116 \pm 8$ pc \\
V mag  & 7.2  \\ 
L' mag & 6.83 \\
L' flux (Jy) & 0.46\\
T$_{eff}$ (K) &10,000 \\
Age &$5\pm 3$ Myr\\
\enddata
\end{deluxetable}

Very recently, \citet{2016arXiv160106560K} presented a HST/STIS image of a new disk component at 65-100 AU around \thisstar, nested inside the two outer belts (see the dashed and plain white contours in Fig.~\ref{fig1}). This optical image seems at first sight to be the scattered light counterpart of the known resolved thermal emission discussed above. We will show later that there is very little overlap, if any, between the thermal emission and the scattered light HST/STIS component reported in \citet{2016arXiv160106560K}. The latter would thus be tracing a fourth disk component and dust population.

We also note two recent independent detections of complex inner disk features from \citet{2016ApJ...819L..26C} and \citet{Perrot:2016bt}. \citet{2016ApJ...819L..26C} reported the detection in the L' band, using Keck/NIRC2 in saturated mode and using angular differential imaging. \citet{Perrot:2016bt} presented Y, J, H and Ks-band images obtained with the second-generation extreme-adaptive optics planet-finder SPHERE at the Very Large Telescope (VLT), using angular differential imaging as well. These detections are, by nature, and due to the disk geometry and complex morphology, affected by observational biases \citep{Milli2012}. Here we present a new high signal-to-noise ratio coronagraphic L'-band image obtained with the near-infrared imager and spectrograph NIRC2 at W.M. Keck observatory.

\section{Observations and data reduction}\label{sec:obs}

We observed \thisstar on June 11, 2015 (see Table~\ref{tbl-1.5}), during the commissioning of the new L'-band vector vortex coronagraph installed in NIRC2 (Serabyn et al.~2016, in press), the near-infrared camera and spectrograph behind the adaptive optics system of the Keck II telescope at W.M. Keck Observatory. The vortex coronagraph is a phase-mask coronagraph enabling high contrast imaging at very small angles close to the diffraction limit of the 10-meter Keck telescope at $3.8 \mu$m ($\simeq 0".08$). The starlight suppression capability of the vortex coronagraph is induced by a $4\pi$ radian phase ramp wrapping around the optical axis. When the coherent adaptively-corrected point spread function (PSF) is centered on the vortex phase singularity, the on-axis starlight is redirected outside the image of the telescope pupil formed downstream from the coronagraph, where it is blocked by means of an undersized diaphragm (the Lyot stop). The vector vortex coronagraph installed in NIRC2 was made from a circularly concentric subwavelength grating etched onto a synthetic diamond substrate \citep[Annular Groove Phase Mask coronagraph or AGPM,][]{2005ApJ...633.1191M,2013A&A...553A..98D}. The mask is nearly identical to its siblings installed inside NACO at the Very Large Telescope \citep{2013A&A...552L..13M, 2013A&A...559L..12A}, and LMIRCam at the Large Binocular Telescope \citep{2014SPIE.9148E..3XD}. See also \citet{Vargas2016} for more details about the AGPM technology and manufacturing process. 
\begin{deluxetable}{cc}
\tabletypesize{\scriptsize}

\tablecaption{Observing log \label{tbl-1.5}}
\tablewidth{0pt}
\tablehead{
\colhead{Properties} & \colhead{value} 
}
\startdata
UT date (dd-mm-yyyy)) & 06-11-2015 \\
UT start time (hh:mm:ss) & 08:10:55 \\
UT end time (hh:mm:ss) & 10:04:44 \\
Discr. Int. Time (s) & 0.2 \\
Coadds & 100 \\
Number of frames & 39\\
Total integration time (s) & 780\\
Plate scale (mas/pix) & 9.942 (``narrow'')\\
Total FoV & $r\simeq 5"$ (vortex mount)\\
Filter &L'\\
Coronagraph &Vortex (AGPM)\\ 
Lyot stop &Inscribed circle\\
Reference PSF &HD\, 144271\\
Median seeing (") &0.5\\ 
Par. angle start-end ($^\circ$) &-24 -- +26\\
Mean airmass &1.1 \\
\enddata
\end{deluxetable}

Observing conditions were very good with optical median seeing $\simeq 0".5$ (estimated from the adaptive optics system telemetry). The adaptive optics system provided excellent correction in the L'-band with Strehl ratio in excess of $85\%$, similar to the image quality provided at shorter wavelengths by more recent extreme adaptive optics systems such as the Gemini Planet Imager and VLT/SPHERE. 

The alignment of the star onto the coronagraph center, a key to high contrast at small angles, was performed using a new method named quadrant analysis of coronagraphic images for tip-tilt sensing \citep[QACITS,][]{Huby2015}. The QACITS pointing control uses NIRC2 focal-plane coronagraphic science images in a closed feedback loop with the Keck adaptive optics tip-tilt mirror (through the real-time update of Shack-Hartman centroid offsets) to guarantee optimal centering at all time (Serabyn et al. 2016, in press). It also significantly improves reproducibility and stability of the centering onto the vortex between targets (E.\ Huby et al., in prep). The typical centering accuracy provided by QACITS is $\simeq 0.025\lambda/D$ rms, or $\simeq 2$ mas rms. We observed a calibrator star, HD\, 144271, an A0 star with similar V and L magnitude to \thisstar. We switched back and forth between target and reference star every 20 minutes. Re-acquisition overheads when switching from the target to the reference and vice-versa were minimized to less than two minutes (including re-centering behind the coronagraph). 
\begin{figure*}[!ht]
  \centering
\includegraphics[width=18.cm]{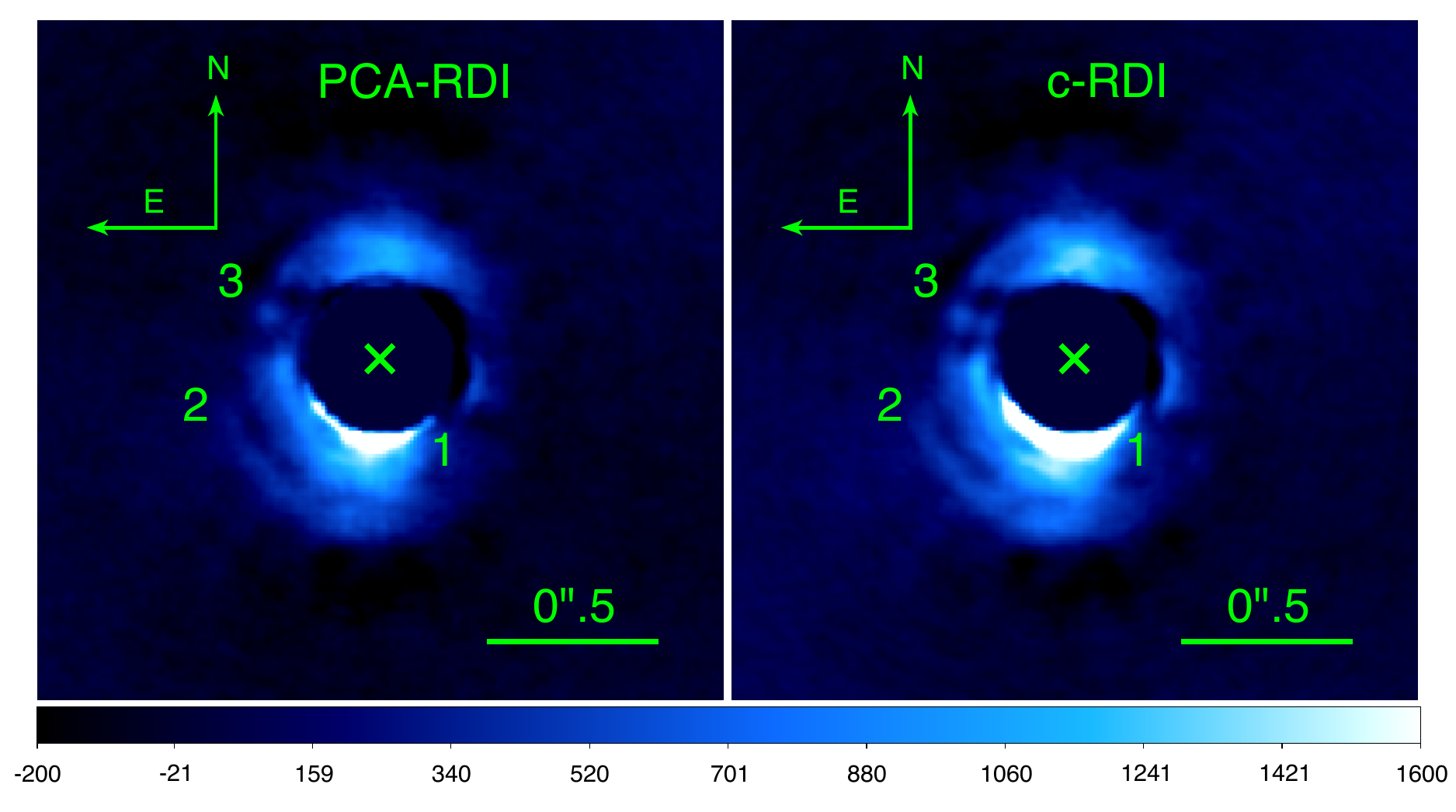}
  \caption{Zoom on the inner disk component with PCA-RDI (left) and c-RDI (right), marking remarkable features. 1: North-South asymmetry, 2: arc-like feature, 3: point-like source seemingly embedded in a gap in the ring. The color scale in both images is linear and in analog to digital units (ADU).\label{fig1b}}
\end{figure*}

The data was reduced by dividing the images by a sky flat field obtained without the vortex phase mask. No sky background image was subtracted as this introduces additional noise unnecessarily. Instead we subtracted a reference PSF constructed by projecting the target images on a subset of the Karhunen-Lo\`eve basis (five first principal components) of the reference star. Reference star differential imaging (RDI) with principal component analysis (PCA-RDI) is very effective in the high Strehl regime provided that the duty cycle between the target and reference star beats speckle decorrelation timescales for a given target contrast \citep{Mawet2009,Serabyn2010,Mawet2011,2012ApJ...755L..28S}. It also is immune to the self-subtraction phenomenon which affects angular and spectral differential imaging (ADI/SDI), especially at small inner working angles. It might however be subject to over-subtraction, which could impact photometric precision \citep{Pueyo2016}. 

To verify and ensure that over-subtraction is indeed minimum, we also used a classical reference star PSF subtraction (c-RDI), by computing the median image of the reference target data cube (we excluded the 10~\% of the reference frames which are the least correlated to the science frames), and removing it  from the target star frames, with a scaling coefficient equal to its projection on the target star frames. This process reveals the inner disk with a similar image quality as when using PCA. Our final reduced images with PCA-RDI and c-RDI are shown in Fig.~\ref{fig1b}, left and right, respectively. Both reduced images are virtually the same, which indicates that the Keck L'-band PSF is very stable. The differences between ADI and RDI can be seen by comparing this image to the ADI-reduced of \citet{2016ApJ...819L..26C}. Our RDI image is not affected by typical ADI self-subtraction biases (akin to high-pass filtering), and so the fainter and more extended disk components are revealed, extending out to $\simeq 0".6$. In particular, our RDI image appears more ring-like, as opposed to two point-like peaks on each side of the star.

\section{Detection of the third inner disk component}\label{sec:det}

We detect the third disk component to the complex debris disk structure around \thisstar. This disk component is the innermost of the three discovered so far in the \thisstar system (Fig.~\ref{fig1} and Fig.~\ref{fig1b}), and is detected from our effective inner working angle of $0".2$ up to $0".6$, corresponding to 23 and 70 AU, respectively. This range also roughly coincides with the resolved thermal emission, so the L'-band image seems to show the short-wavelength counterpart of the extended thermal feature detected in the mid-infrared (Q and N bands) by Keck/OSCIR, Keck/MIRLIN, and VISIR more than a decade ago \citep{Fisher2000,Marsh2002,2014A&A...561A..50T}.

The vortex coronagraph has a theoretical inner working angle of $0.9 \lambda/D \simeq 0".07$, with the wavelength $\lambda = 3.8$ $\mu$m and the telescope diameter $D\simeq 10$ m. Due to the central obscuration, segmented pupil geometry, and Lyot stop size, the Keck/NIRC2 vortex has a measured inner working angle of $1.6 \lambda/D \simeq 0".125$. The --effective-- inner working angle is affected by the residuals due to imperfect reference subtraction, and is for this particular data set closer to $\simeq 2\lambda/D$. 
\begin{figure}[!ht]
  \centering
\includegraphics[width=9.5cm]{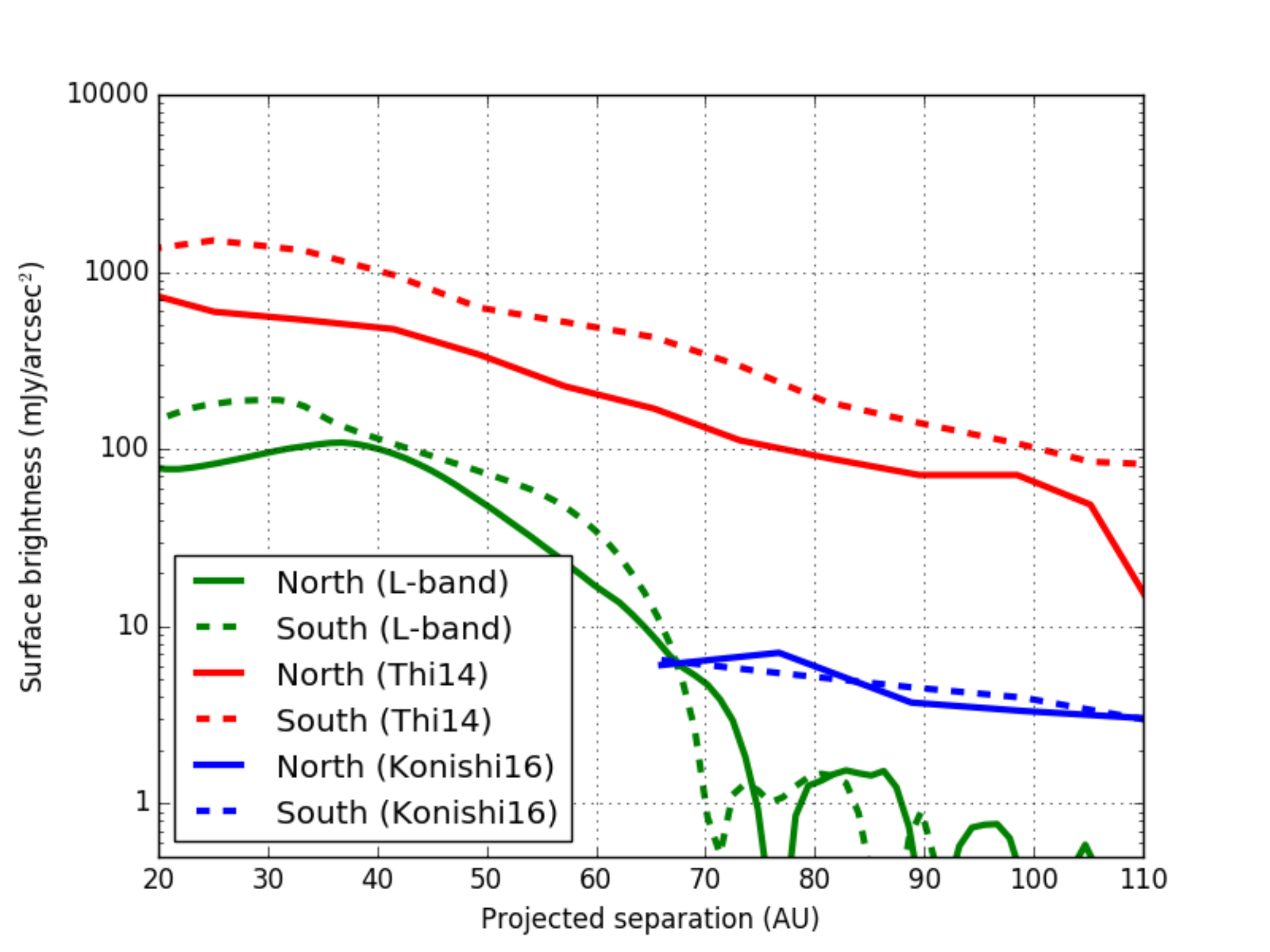}
  \caption{Surface brightness profiles (mJy/arcsec$^2$) of the disk extracted along its major axis, and using aperture photometry with a radius $r=\lambda/D\approx 0".08$, towards the North (plain curve) and South (dashed curve), respectively. The green curve is our L'-band data. We overplot the published VISIR $8.6 \mu$m PAH profile from \citet{2014A&A...561A..50T} in red, and the published HST/STIS optical profile from \citet{2016arXiv160106560K} in blue. \label{fig2}}
\end{figure}

For typical dust grain composition and size in debris disks, the equilibrium temperature at the separations probed by our observations is $\simeq 100$ K, which is warmer than typical cold Kuiper belts found by Spitzer or Herschel at $\leq 50$K, but not as warm as asteroid belt dust ($\sim 150$~K) or zodiacal dust ($\sim 300$~K) in the solar system. The $3.8 \mu$m thermal emission of dust disk models peaks at a radius much smaller than observed, pointing to a scattered light origin for the disk component detected at L'. Moreover, the brightness profiles extracted over the disk major axis (using aperture photometry with a radius $r=\lambda/D\approx 0".08$) presented in Fig.~\ref{fig2} show a steep $r^{-4}-r^{-5}$ power law between 23 and 70 AU, compared to the $\simeq r^{-2.5}$ of the VISIR $8.6 \mu$m PAH image of \citet{2014A&A...561A..50T}. The slope difference implies that the disk should become redder at larger separations. However, PAH emission is an unreliable tracer of dust, and we chose not to make direct comparisons between these images.

We note that both the PCA-RDI and c-RDI images show three remarkable features (noted 1, 2, and 3 in Fig.~\ref{fig1b}), although we recognize that they could be artifacts resulting from imperfect coronagraph centering. Indeed, the data was taken during the vortex coronagraph commissioning run and the pointing was thus at times sub-optimal (e.g., the QACITS set point was likely not fully consistent between the target and reference, inducing differential structures mostly within $0".2$ but also potentially slight changes in the diffraction pattern beyond). 

The first feature is a noticeable N-S asymmetry (the South side being about twice as bright as the North side, see Fig.~\ref{fig2}) that could also be due to pointing errors on the coronagraph. However, the feature is also seen in \citet{2016ApJ...819L..26C}, and follows the N-S asymmetry of the ring ``R1'' identified  in the SPHERE images at shorter wavelengths by \citet{Perrot:2016bt}, which reveals a fine ring traced to the South, almost invisible to the North. 

The second interesting feature is an arc-like structure to the South East at a radius $r\simeq 0".6$. A reminiscent feature is reported in \citet{2016ApJ...819L..26C}, and in the near-infrared SPHERE images. In the SPHERE images, it is identified as ring ``R2'' \citep{Perrot:2016bt}. It is also seen in the optical HST/STIS image, and identified as a ``arc-like'' feature \citep{2016arXiv160106560K}.

The third feature is a point-like structure to the East (FWHM $\approx 1 \lambda/D$). The projected angular separation is $0".330 \pm 0.010$, corresponding to a projected physical separation of $\simeq 38$ AU and position angle of $70^\circ \pm 2$. Using aperture photometry with an aperture radius $r=\lambda/D$, we find that this point-like feature is marginally detected at a signal-to-noise ratio of $\simeq 2$, and $\simeq 9$ mag fainter than the central star in the L' band (L'=$15.8^{+0.75}_{-0.45}$). The low-significance detection could also be a dust clump embedded in the inner disk component. We also note that, while other point-like feautres are seen in their images, our particular detection is not reported in the saturated Keck/NIRC2 data set of \citet{2016ApJ...819L..26C} nor in the SPHERE near-infrared images of \citet{Perrot:2016bt}.

\section{Disk modeling}

Besides characterizing the morphology of the inner disk around \thisstar, this new image in the L' band allows us to put some constraints on the dust composition of the system, as we will discuss in this section. Although the model proposed by \citet{2014A&A...561A..50T} provides a good fit to the 8.6~$\mu$m PAH emission image and to the spectral energy distribution (SED), it fails at reproducing properly our L'-band image of the inner disk component, which is expected. Without an image of the dust continuum, there is indeed a strong degeneracy between the disk morphology and composition parameters, especially for a complex multi-belt system such as \thisstar.

To characterize the morphology and composition of the inner disk component, we proceeded in two steps with complementary modeling tools.
First, we constrained the geometry of the inner disk using the GRaTer ray-tracing code \citep{Augereau1999a,2012A&A...539A..17L}, which creates scattered-light images of optically thin debris disk models assuming simple anisotropic scattering \citep{Henyey1941}. These simplified models allow us to estimate offsets of the disk with respect to the star and compare them with already published measurements. 
We then used the Monte Carlo radiative transfer code MCFOST \citep{Pinte2006,Pinte2009} to analyze the morphology and composition of the complete dust system altogether by fitting models to both our L'-band image and the system SED.

\subsection{Modeled datasets}

We fit our models to the L'-band image of the disk obtained after a classical reference star PSF subtraction (c-RDI, see Sect.~\ref{sec:obs}). When computing goodness of fit estimators, we masked out the pixels outside of an ellipse centered on the disk of semi-minor and semi-major axes 0\farcs57 and 0\farcs66 respectively, to consider only the area where the disk is detected as well as a thin $\sim 1\lambda/D$-wide ring around the disk where it is not detected. We also masked the inner part of the image out, within a radius of 0\farcs22, leaving a total of $N_{im}=10286$ pixels in the fit process.

To estimate the noise map used in the least-square fits, we adapted the procedure described in \citet{2016ApJ...817L...2C}, and proceeded as follows: (1) we estimated the noise on a single NIRC2 exposure by reducing the 23 images of the reference star with the same method as used for \thisstar, then computing the ``temporal'' standard deviation for each pixel through these reference images; (2) we then estimated the noise on the final image of \thisstar (which combines 36 exposures rotated to North up) by adding in quadrature 36 replicas of the single exposure noise map, de-rotated by the same parallactic angles as for the \thisstar frames. This method allows us to estimate the local noise in the image (including smearing of speckle noise due to the derotation and combination of images over a large parallactic angle range), even in regions with extended astrophysics signal, as opposed to classical radial noise curves estimated from spatial noise in the science image itself. 

As a sanity check, we also estimated the background noise based on the photon counts in our image, and we found a similar noise level as in our noise map. This demonstrates  that our reduced image is mostly dominated by background noise, except close to the star  where the noise source is a combination of speckle noise and background noise, which is accurately captured by our temporal noise calculation.

In section \ref{sec:mcfostgrid} we additionally used the system SED to further constrain the dust characteristics in \thisstar. We used  the photometric data at wavelengths longer than 5~$\mu$m listed in \citet{2014A&A...561A..50T}, binned to have as many photometric points per decade ($N_{SED}=19$ data points). To account for under-estimated error bars, cross-calibration systematics between instruments, and natural variability, we set all the uncertainties to 10\% in the fitting process. The SED photometric measurements are presented in Fig.~\ref{fig:best model}. The stellar photometry was corrected from reddening based on \citet{Cardelli1989} and using \citet{ODonnell1994} updated constants for the near-UV.

\subsection{Geometry of the inner disk}

To constrain the most relevant morphological parameters of the inner disk component, we created a grid of $\sim1300$ models with the GRaTer code, using the inclination, position angle, parent dust belt radius, disk offsets, and degree of forward scattering as free parameters. We used a Gaussian profile for the vertical dust density distribution, with a linear vertical scale height of an opening angle of 5\% \citep{2014A&A...561A..50T}. We simulated the surface density distribution of the disk with a combination of two radial power laws, with slopes arbitrarily set to $\alpha_{in}=3$ and $\alpha_{out}=-5$ respectively inward and outward from the parent belt radius. As \citet{2016ApJ...819L..26C}, we found that these parameters are not constrained by our data. The total flux of the normalized models are adjusted to the data using a linear regression with the observed disk.
\begin{deluxetable}{cc}
\tabletypesize{\scriptsize}
\tablecaption{Geometrical properties of the innermost disk component around \thisstar \label{tbl-2}}
\tablewidth{0pt}
\tablehead{
\colhead{Properties}  & \colhead{model} 
}
\startdata
Position angle of major axis ($^\circ$) &$ -11 \pm 8$ \\
Parent belt location $r_0$ (AU) &$39 \pm 4$  \\ 
Inclination ($^\circ$)  &$ 53\pm 6$ \\
Disk offset in x $x_c$ (AU)   & $-2\pm7$\\ 
Disk offset in y $y_c$ (AU)  & $0\pm4$\\
Coeff.~of forward scatt.~$g$ &$0.1\pm0.1 $ \\
\enddata
\end{deluxetable}

The parameters of the best model are reported in Table~\ref{tbl-2}, and correspond to an almost isotropic disk (Henyey-Greenstein asymmetric scattering factor $g=0.1$), with a parent belt radius of 39~AU ($\sim0.34\arcsec$) inclined by $53\degr$ with a position angle of $-11\degr$ North to East. These values are consistent with the corresponding parameters for the two outer rings, both regarding the geometrical parameters \citep{2015MNRAS.450.4446B} and the anisotropy of the dust grains \citep{1999ApJ...525L..53W}. We note that our best-fit geometry differs from the characteristics estimated by \citet{2016ApJ...819L..26C}, although using the same modeling code. For instance, we estimate the disk offsets from the star to be $x_c=-2\pm7$ AU and $y_c=0\pm4$ AU in the detector frame, which are consistent with a belt centered on the star position. Our $x_c=-2\pm7$ AU value significantly differs from the $7.9$ AU offset found by \citet{2016ApJ...819L..26C}. 

One possible explanation for the discrepancy is that \citet{2016ApJ...819L..26C} used a fixed $-1.2\degr$ position angle value, and their best fit $56\degr$ inclination hits the lower limit of their model grid. However, it is worth noting that \citet{2016ApJ...818..150M} did not find any significant offset for the 2 outermost rings in their re-analysis of Gemini/NICI and HST/NICMOS data, in disagreement with \citet{2015MNRAS.450.4446B}.

\subsection{Detailed morphology and chemical composition of the inner disk\label{sec:mcfostgrid}}

To put constraints on the dust composition in the \thisstar disk system, we fit our L'-band image and the system SED to models generated with the radiative transfer code MCFOST. We computed a grid of 194400 models with $N_{par}=10$ varying parameters to describe the inner disk morphology and composition:  the inner and outer radius of the disk $R_{in}$ and $R_{out}$, the vertical scale height $H_0$ at 100~AU from the star, the surface density exponent $\alpha$, the flaring exponent $\beta$, the dust mass $M_{dust}$, the minimum and maximum grain sizes $a_{min}$ and $a_{max}$, the grain porosity $p$, and the dust composition. Neglecting pressure drag from the gas, we fixed the dust size distribution to the standard collisional cascade power law with exponent $-3.5$ from \citet{Dohnanyi69}. We adopted homogeneous spherical dust grains, and used the Mie theory to derive their optical properties: the scattering and absorption opacities, as well as the scattering phase functions. We used 4 different dust species in our grid of models: pure olivine grains \citep{Dorschner1995} as used by \citet{2014A&A...561A..50T} to model the three dust belts in \thisstar, and three different mixtures of amorphous silicates \citep{Draine1984}, amorphous carbon \citep{Rouleau1991}, and water-dominated ice \citep{Li1998}. The three mixtures all have a different dominant component, with the respective proportions 1:2:3 (water-ice dominant), 2:3:1 (carbon dominant), and 3:1:2 (silicate dominant). The parameter values simulated in the model grid are described in table~\ref{tab:grids}. The morphology and composition of the two outer belts and the PAH content in the inner disk were kept fixed to the values adopted by the best fit derived in \citet{2014A&A...561A..50T}.

\begin{deluxetable*}{lcccl|ccc}
\tabletypesize{\scriptsize}
\tablecaption{Parameters probed in our grid of 194400 MCFOST models, and best fits to the SED and L'-band data. \label{tab:grids}}
\tablehead{
 \colhead{Parameters}	 & \colhead{Min. value}	 & \colhead{Max. value}		 & \colhead{Sampling}& \colhead{$N_{sample}$}	& \colhead{Best SED}	 & \colhead{Best image} 	 & \colhead{\textbf{Best comb.}}}
\startdata
$M_{dust}$ ($M_\sun$)   & 0.2e-6 	&3.2e-6	    &log.		&5 		&0.2e-6		&0.2e-6\tablenotemark{a}		&\textbf{0.2e-6}	\\
$a_{min}$ ($\mu$m)		&0.1		&10			&log.		&5		&0.5		&10    		&\textbf{0.1}	\\
$a_{max}$ ($\mu$m)		&1000	    &10000		&log.		&3	    &10000		&5000  		&\textbf{10000}	\\
$R_{in}$ (AU)			&20	        &40		    &lin.		&3		& 20    	&30			&\textbf{20}	\\
$R_{out}$ (AU)			&70	        &110		&lin.		&3		&110		&70    		&\textbf{90}	\\
$H_{0}$ (AU)			&5	        &20			&log.		&3		&5		 	&10    		&\textbf{20}	\\
$\alpha$				&-2.5	&-0.5	    	&lin.		&3   	&-2.5	 	&-0.5   		&\textbf{-0.5}	\\
$\beta$				    &1.00	&1.25	    	&-		    &2 	 	&1.25\tablenotemark{a} 		&1.25\tablenotemark{a}   		&\textbf{1.25}\tablenotemark{a}	\\
$p$ (\%)				&0.0        &0.9	    &lin.		&4	    &0.0		&0.3			&\textbf{0.0}	\\
Dominant species 		&\multicolumn{3}{c}{Olivine, Silicate, Carbon, Water ice}&4&Olivine &Olivine	&\textbf{Olivine}\\
\hline
$\chi^2_{SED}$		&-&-&-&-						    			&\textbf{32}	&61		    &\textbf{35}	\\
$\chi^2_{im}$		&-&-&-&-									    &22 		    &\textbf{10}	&\textbf{14}	\\
$\chi^2_{comb}$	&-&-&-&-									        &54	    	    &71		    &\textbf{49}	\\
\enddata
\tablenotetext{a}{Not well constrained by our Bayesian analysis. See Fig.~\ref{fig: proba density}.}
\end{deluxetable*}

MCFOST generates SED data points at prescribed wavelengths, as well as synthetic images. The model images at 3.8$\mu$m are convolved with the L'-band Keck PSF measured off-axis during the coronagraphic acquisition sequence. We note that the absolute flux of disks in scattered light is poorly predicted by radiative transfer algorithms, due to poor observational constraints on the dust scattering phase functions at small scattering angles \citep{2015ApJ...811...67H}. Combined fits of unscaled model images to scattered light images and SEDs usually lead to aberrant results, as we experienced in this study and was also reported by \citet{2006ApJ...650..414S} and \citet{2015A&A...577A..57M}. To address this known problem, we used a free scaling coefficient to adjust the flux of the model images to the flux of the observed disk, which is equivalent to relaxing the constraint on the dust mass in the image \citep{2012A&A...539A..17L}.
\begin{figure*}[!ht]
   \centering
 \includegraphics[width=17cm]{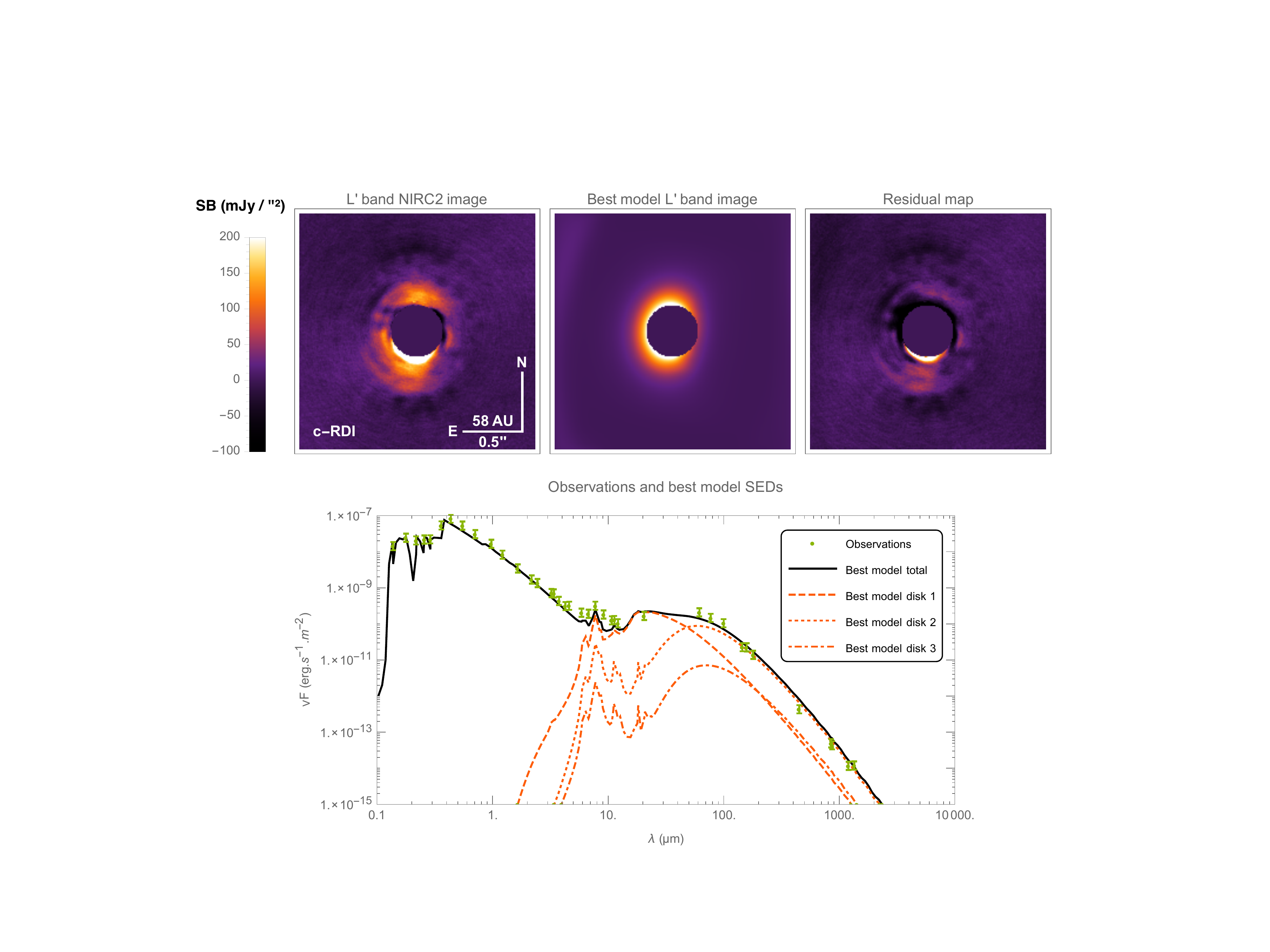}
   \caption{Most likely model fitting both our L'-band image of \thisstar innermost disk and its SED altogether. The top row shows the L'-band image of \thisstar obtained with the Keck-NIRC2 vortex coronagraph after classical reference star PSF subtraction (left), the 3.8~$\mu$m image of the best model (see Table~\ref{tab:grids}) (middle), and the residual map obtained by subtracting the image of the best model from the L'-band image. All three images are shown with the same field of view and the same linear scaling. The bottom row shows the measured SED of the system (green points), and the SED of the best model (black line). The red lines shows the individual contributions of the three dust belts in the best model. \label{fig:best model}}
 \end{figure*}

To assess the goodness of fit of our models to the data, we computed three $\chi^2$ estimators for each model: a reduced chi square on the SED $\chi^2_{SED}$ (9 degrees of freedom), a reduced chi square on the L'-band image $\chi^2_{im}$ (10276 degrees of freedom), and a combined estimator defined by:
\begin{equation}
\chi^2_{comb} = \chi^2_{SED}+ \chi^2_{im}.
\end{equation}

The $\chi^2_{im}$ estimators are computed from the sum of the squared pixel-to-pixel differences between the model image and the observed disk image weighted by the estimated noise map, and are normalized by the number degrees of freedom $\nu_{im}=N_{im}-N_{par}$, where $N_{im}=10286$ is the number of pixels in the image, and $N_{par}$ the number of model parameters (here 10). The $\chi^2_{SED}$ estimators are computed from the sum of the squared point-by-point differences between the model SED and the observed SED weighted by the data error bars, and are normalized by the number degrees of freedom $\nu_{SED}=N_{SED}-N_{par}$, where $N_{SED}=19$ is the number of data points in the SED, and $N_{par}=10$.

Table~\ref{tab:grids} summarizes the parameters varied in our modeling, as well as the parameters and chi square values of the models that best fit the SED, the L'-band image, and the two datasets altogether. The SED and image of the model that best fit the observed SED and NIRC2 L'-band image of \thisstar simultaneously are also presented in Fig.~\ref{fig:best model}. This best model provides an excellent compromise that adequately fits both the measured system SED and the NIRC2 L'-band image of the inner disk, as demonstrated by its $\chi^2_{SED}$ and $\chi^2_{im}$ very close to their respective best values. The main differences between the best model and our data are the non-centro symmetric features in the L'-band image discussed in Sec.~\ref{sec:det}, which cannot be properly fit by our symmetric disk model: the N-S brightness asymmetry and the arc at the Southern extremity of the disk. 
 \begin{figure*}[!ht]
   \centering
 \includegraphics[width=17cm]{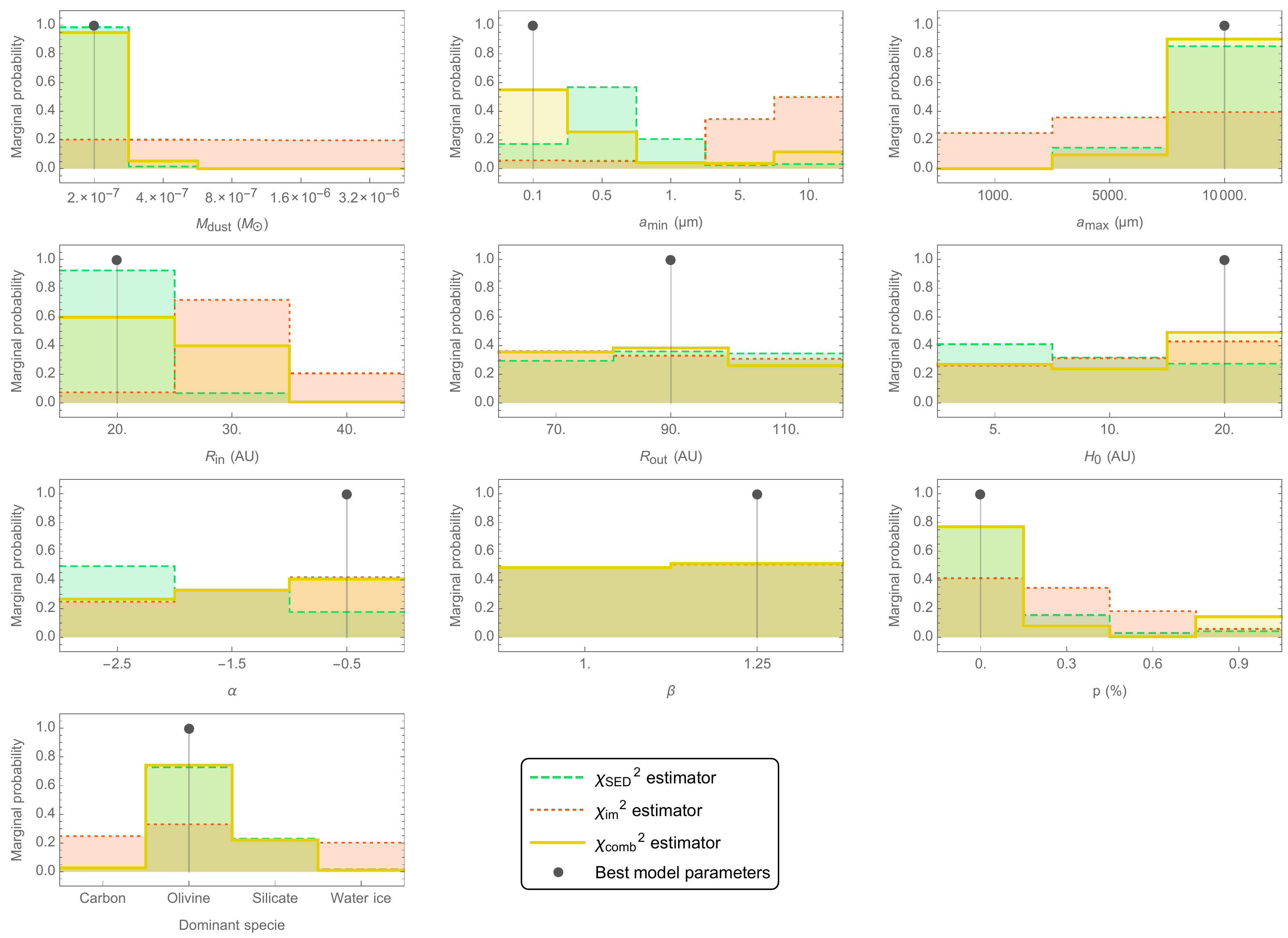}
   \caption{Relative probability distributions of each parameter given the data and the grid of models, with respect to the SED (dashed green), to the L'-band image (dotted red), and to both datasets simultaneously (solid yellow). The black dots mark the parameter values of the best model. See text for details. \label{fig: proba density}}
 \end{figure*}

 We used the Bayesian formalism to determine the validity range of each parameter around their best values \citep[see e.g.][]{2008A&A...489..633P,2015A&A...577A..57M}. Under the assumption that the noise affecting our independent datasets follows a Gaussian distribution, the probability of occurrence of a given model given our data is related to the chi square value by the formula:
 \begin{equation}
 P \propto \exp \left(-\frac{\chi^2}{2}\right).
 \end{equation}
 By normalizing the probabilities of our models by the sum of the probabilities over the entire grid, we computed the relative probability of each model, given our data and the grid of models. We present in Fig.~\ref{fig: proba density} the probability distributions obtained by marginalizing the relative probabilities over each parameter value, for each of the chi-square estimators. 
 
 As expected, this analysis shows that the SED alone poorly constrains the geometry of the system, and the L'-band image alone poorly constrains the dust properties. The probability distribution analysis presented in Fig.~\ref{fig: proba density} indicates that the dust mass in the disk is not well constrained by our image (flat probability distribution), but is sharply constrained by the SED. We found a most-probable value of 2.0e-7~$M_\sun$, consistent with the one constrained by \citet{2014A&A...561A..50T}, and which confirms that the inner disk is much less massive than the outer rings. The grain size ranges from $0.1~\mu$m to 10 mm, also mostly constrained by the SED, and consistent with values found in \citet{2014A&A...561A..50T}. We note that the most probable value for $a_{max}$ reaches the highest value allowed in our grid, and is thus probably a lower limit. This demonstrates that the disk is composed of large grains, typical of an evolved debris disk. Given the large grain size at the tail of the distribution ($a_{max}=10$~mm), most of the mass in the $M_{dust}$ parameter is contained in the millimeter-size grains, as also noted by \citet{2014A&A...561A..50T}. Our L'-band image mainly helps constraining the geometrical parameters of the disk, with most likely values of 20~AU for the scale height $H_0$ at a radius of  100~AU from the star, of -0.5 for the surface density exponent $\alpha$ (the disks spreads outwards due to the abundance of small grains which must have eccentric orbits), 90~AU for the outer radius of the disk $R_{out}$ consistent with thermal infrared data \citep{Fisher2000,Marsh2002}, and 20~AU for the inner disk radius $R_{in}$, although all chi-square estimators suggest that this is an upper limit, given that we do not detect the inner boundary of the disk in our image. We note that none of our datasets enable to constrain the flaring coefficient of the disk. Finally,  all three of our goodness-of-fit estimators point with a high probability toward a composition of non-porous olivine grains for the inner disk of the system, as was assumed for the outer rings by \citet{2014A&A...561A..50T}. 

\section{Giant planet detection limits\label{sec:detlims}}

The sensitivity of the Keck/NIRC2 vortex coronagraph to detect giant exoplanet is very good. In Sect.~\ref{sec:det}, we reported the detection of a putative and very faint point-like source. If real and associated, the source would hypothetically correspond to a $10\, M_J$ companion, assuming the BT-SETTL evolutionary model from \citet{Allard2014}.

\begin{figure}[!ht]
  \centering
\includegraphics[width=9cm]{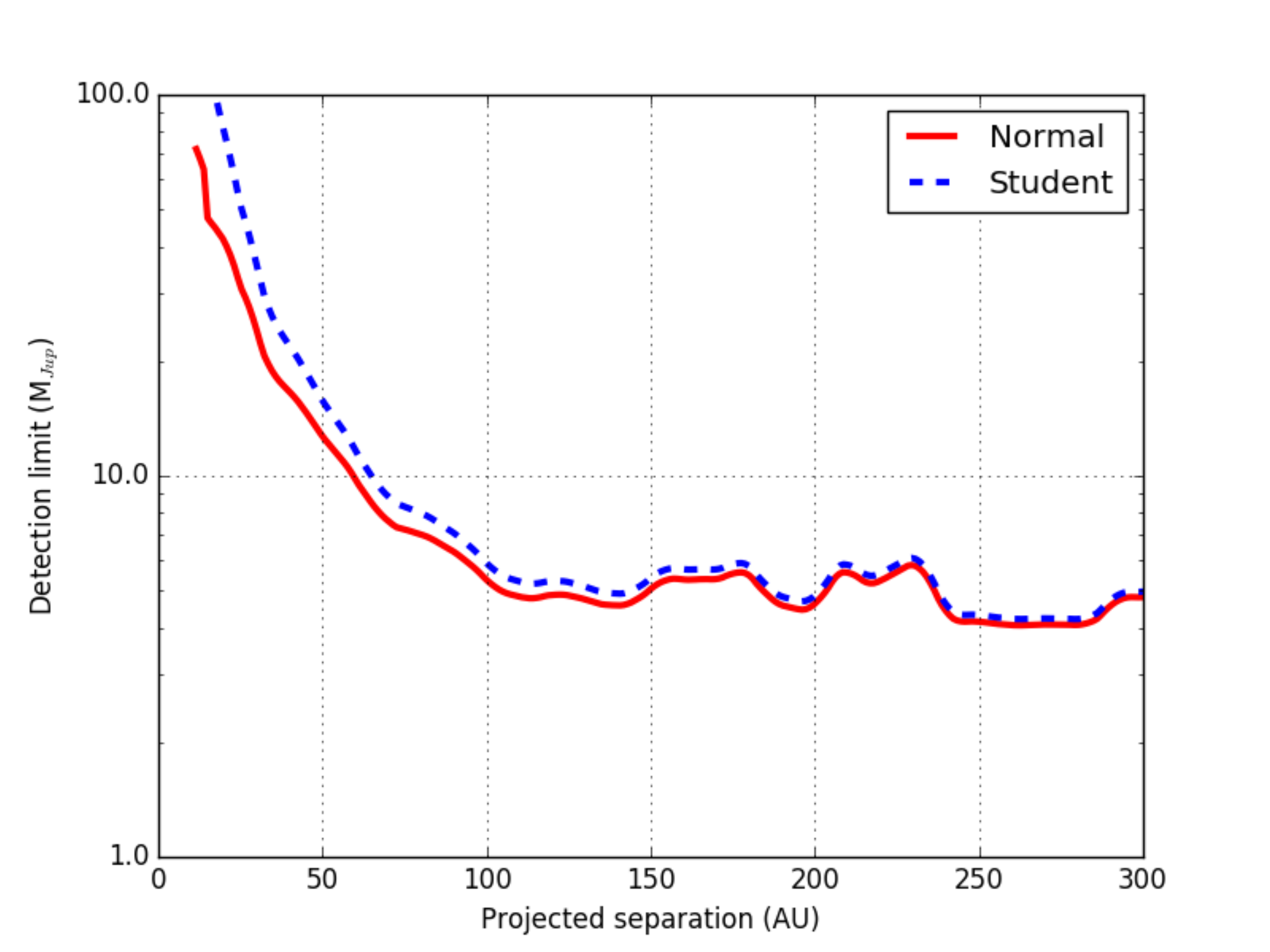}
  \caption{Detection limits ($5\sigma$) around \thisstar in units of Jupiter mass. The inner working distance is about 10 AU. The first 100 AUs are contaminated by the disk scattered light. Far from the star and outside the inner disk, our detection limit is close to 4-5 $M_{Jup}$. We use the BT-SETTL evolutionary model of \citet{Allard2014}. The red plain curve shows the traditional $5\sigma$ detection limits, while the dashed blue curve shows the same detection limits corrected for small sample statistics effect using the Student t-distribution as in \citet{Mawet2014}. \label{fig4}}
\end{figure}

Ignoring the -- only -- candidate point source present in the data and described in Sect.~\ref{sec:det}, we computed our detection limits in and beyond the region of the image covered by the extended disk, using RDI and principal component analysis (PCA) as described in Sect.~\ref{sec:obs}. Our reference PSF was constructed using five principal components. Our $5\sigma$ sensitivity is uniform and close to $\simeq 5 M_J$ (see Fig.~\ref{fig4}), assuming the BT-SETTL evolutionary model from \citet{Allard2014}. Following \citet{Mawet2014}, we accounted for small sample statistics penalties at small inner working angles using the Student t-distribution.

\section{Discussion\label{sec:discussion}}

The complex circumstellar environment of five Myr-old \thisstar is undergoing dramatic changes, evolving from the proto-planetary stage to debris disk. Gas and dust are being processed by physical interactions related to stellar evolution and planetary formation: grain growth, dust settling, gas photo-evaporation, Poynting-Robertson drag and blowout by radiation pressure. This transition is occurring over the timescales of giant planet formation, with its minutiae determining the architecture of the future planetary system. Our detailed analysis sheds new light on the inner disk component around \thisstar. Our lower-limit on the maximum dust grain size is 10 mm, which is reminiscent of older debris disk, and showing that significant grain growth has already occurred for this relatively young system. On the other hand, the minimum grain size constrained by our analysis, $0.1 \mu$m, is well below the blowout size from radiation pressure for \thisstar ($\simeq 4 \mu$m), suggesting a very active disk where collisions are replenishing the reservoir of small grains constantly. The inner disk has some properties of a young protoplanetary disks and it may not have reached collisional equilibrium yet. Moreover, the presence of small grains being blown out is consistent with our finding of -0.5 for the surface density exponent $\alpha$, which indicates that the disks spreads outwards due to the abundance of small grains on eccentric orbits.  

Contrary to the conclusions of \citet{Li2003}, our best fit to the SED and L'-band image shows with a high confidence level that the inner disk is composed of a much simpler grain structure and composition. Confirming the assumptions adopted in the disk model used in \citet{2014A&A...561A..50T}, we find that non-porous olivine grains have a much higher marginal probability than any of the porous carbonaceous, silicate and water ice mixtures considered here. Our data and analysis thus indicates that the inner disk seems indeed devoid of water ice. The fact that we find non-porous grains is also suggestive of asteroidal grains, as opposed to fluffy icy cometary aggregates found, for instance, in older debris disks such as HD\,181327 \citep{2012A&A...539A..17L}. The composition of the inner disk does not differ significantly from the outer disks, as presented in \citet{2014A&A...561A..50T}, suggesting a common origin.

It is interesting to compare our L'-band data and analysis with the HST/STIS detection presented by \citet{2016arXiv160106560K}, and tempting to assume that both detections overlap and see the same inner disk component. In reality, there is no significant overlap between our L'-band data and the optical image, as the disk is not detected beyond 70~AU in our L'-band image, and \citet{2016arXiv160106560K}'s HST/STIS detection ranges from 65 AU to 100 AU along the disk major axis. The L'-band data thus probes much smaller projected separations, and a different dust population with larger grains than seen in the STIS visible image. As our MCFOST model also generates optical images, we decided to compare our best-fit model synthetic image to the HST/STIS data. We found that the predicted flux is brighter than the STIS image. Two reasons can be invoked to explain the discrepancy: the phase function chromatic behaviour of our dust particles is not well described by the Mie theory as discussed in previous studies \citep{2015ApJ...811...67H,2006ApJ...650..414S,2015A&A...577A..57M}, and/or the dust population seen at optical wavelengths is different from the dust population seen in the L' band. 

The intermediate near-infrared wavelengths probed by the SPHERE data presented in \citet{Perrot:2016bt} offers an interesting perspective. The disk morphology in the SPHERE images shows a broken, narrow ring with multiple point sources, none of which are superimposed with our data. These image features are likely the consequence of the high-pass filtering characteristics of ADI, which our RDI reduction prevents \citep{Milli2012}. Nevertheless, the broken ring structure resolved in the SPHERE data sits right at the transition between our L'-band image and the optical HST/STIS image. The prominent feature in the near-infrared images seemingly marks a contrasted zone in the surface brightness, reminiscent of a break in dust population characteristics. It is thus likely that the optical image probes a tenuous population of very small grains blown out of the inner disk by stellar radiative pressure, as is common at these spatial scales for disk around the same age and evolutionary stage \citep{2014A&A...566A..91M,2015A&A...577A..57M}. The current picture of the \thisstar inner system shows a very active inner component up to $\simeq 70$ AU, where collisions dominate, creating very small grains, and an outward progression in small dust grain location from the L'-band to the H-band (VLT/SPHERE image) to the visible (HST/STIS image), likely indicative of dust blowout.

Massive companions orbiting exterior to and inclined to a circumstellar disk have recently been shown to excite the disk particle eccentricities and inclinations via the Kozai-Lidov mechanism, producing dust \citep{Nesvold2016}. This mechanism has also been shown by the same author to shepherd rings, providing an alternative scenario to the classical eccentric planet orbiting interior to the disk stirring the larger bodies in the belt, and produce dust via inter-particle collisions. It is possible that the set of concentric, heavily disturbed rings of \thisstar are, at least in part, the results of complex dynamical secular interactions with outer companions. We also note that the spiral-like features seen on multiple scales around \thisstar can also be caused by the presence of interior or exterior companions (including the two M-dwarf companions) launching density waves \citep{Dong2015,Dong2016}. \citet{Weinberger2000} argues that the two other companion stars (types M2 and M4) both around 900 AU away in projected separation are associated to \thisstar, forming a triple system. However, \citet{Reche2009} demonstrated that the radial velocity measurements of \citet{Weinberger2000} are likely wrong, and used the radial velocity from the $CO$ line instead, obtaining a more accurate result. The new measurement indicates that the companion pair likely flew by. The flyby scenario, while attractive, still cannot explain the gaps between the two outer rings \citep{Ardila2005}.\\

\section{Conclusion}

We have presented a new L'-band coronagraphic image of the third disk component around \thisstar. The image was obtained during the commissioning of the L'-band vortex coronagraph installed in Keck/NIRC2. A simple disk model was fitted to the data yielding a parent body belt location at $r=39$ AU. No significant offsets between the disk and the star were observed.

A more complex model generated with the radiative transfer code MCFOST was used to fit both the SED and the new L'-band image and constrain the dust properties and disk morphology, revealing that the disk is composed of non-porous olivine dust grains, extending from below 20~AU and up to 90 AU. We found a total dust mass of 2e-7 $M_\sun$ for grain sizes ranging from 0.1~$\mu$m to 10~mm. The gas mass is therefore 2e-5 $M_\sun$, consistent with \citet{2014A&A...561A..50T}. This population of grains is interior and differs from the one probed in the optical by HST/STIS \citep{2016arXiv160106560K}. Interestingly, the ring structure detected in the near-infrared by VLT/SPHERE sits right at the transition between the inner disk component dust population probed by our data, and the tenuous small grain population likely blown out of the inner disk and seen in the HST/STIS image. 

Our data and multi-wavelength analysis constrained the dust composition of the newly discovered inner disk component of \thisstar for the first time. Our analysis reveals that this five-Myr old disk already agglomerated micron-sized olivine dust grains, similar to evolved debris disks. Studying both the composition of dust grains and their distribution in young transitional disks is key to understanding the last stage of the formation of rocky and giant planets.

\acknowledgments
Based on observations made at the W.M. Keck Observatory, which is operated as a scientific partnership among the California Institute of Technology, the University of California and the National Aeronautics and Space Administration. The Observatory was made possible by the generous financial support of the W.M. Keck Foundation.  We would like to acknowledge J.-C. Augereau for the development and the sharing of the GRaTer disk modeling tool. Support for this work was provided by NASA through Hubble Fellowship grant \#HST-HF2-51355.001-A awarded by the Space Telescope Science Institute, which is operated by the Association of Universities for Research in Astronomy, Inc., for NASA, under contract NAS5-26555.

\facility{W.M. Keck Observatory, Keck II, NIRC2}

\software{GRaTer \citep{Augereau1999a,2012A&A...539A..17L},
		 MCFOST \citep{Pinte2006,Pinte2009}}

\bibliography{hd141_astroph}

\end{document}